\RequirePackage[mathlines]{lineno}
\documentclass[prd,twocolumn,showpacs,amsmath,amssymb]{revtex4-1}
\usepackage{overpic,graphicx}
\usepackage{dcolumn}
\usepackage{bm}
\usepackage{rotating}
\usepackage{subfigure}
\usepackage{color}
\usepackage{epstopdf}
\definecolor{gongnvlan}{rgb}{0.15,0.39,0.46}
\begin{document}
\title{\bf \boldmath Search for the decay $D_{s}^{+} \to p\bar{p}e^{+}\nu_{e}$ }
%   usage: Please add "\input{BESIII_authors}" after \title{...}
%   to your latex file
\author{
M.~Ablikim$^{1}$, M.~N.~Achasov$^{10,e}$, P.~Adlarson$^{63}$, S. ~Ahmed$^{15}$, M.~Albrecht$^{4}$, A.~Amoroso$^{62A,62C}$, Q.~An$^{59,47}$, ~Anita$^{21}$, Y.~Bai$^{46}$, O.~Bakina$^{28}$
% ...
\\
\vspace{0.2cm}
(BESIII Collaboration)\\
\vspace{0.2cm} {\it
$^{1}$ Institute of High Energy Physics, Beijing 100049, People's Republic of China\\
% ...
$^{4}$ Bochum Ruhr-University, D-44780 Bochum, Germany\\
% ...
$^{10}$ G.I. Budker Institute of Nuclear Physics SB RAS (BINP), Novosibirsk 630090, Russia\\
% ...
$^{15}$ Helmholtz Institute Mainz, Johann-Joachim-Becher-Weg 45, D-55099 Mainz, Germany\\
% ...
$^{21}$ Indian Institute of Technology Madras, Chennai 600036, India\\
% ...
$^{28}$ Joint Institute for Nuclear Research, 141980 Dubna, Moscow region, Russia\\
% ...
$^{46}$ Southeast University, Nanjing 211100, People's Republic of China\\
$^{47}$ State Key Laboratory of Particle Detection and Electronics, Beijing 100049, Hefei 230026, People's Republic of China\\
% ...
$^{59}$ University of Science and Technology of China, Hefei 230026, People's Republic of China\\
% ...
$^{62}$ (A)University of Turin, I-10125, Turin, Italy; (B)University of Eastern Piedmont, I-15121, Alessandria, Italy; (C)INFN, I-10125, Turin, Italy\\
$^{63}$ Uppsala University, Box 516, SE-75120 Uppsala, Sweden\\
% ...
$^{e}$ Also at the Novosibirsk State University, Novosibirsk, 630090, Russia\\
% ...
}
}

\begin{abstract}
    Using a $3.19~\mathrm{fb}^{-1}$ data sample collected at the $\sqrt{s}
    ~=~4.178$ GeV with the BESIII detector, we search for the rare decay
    $D_{s}^{+} \rightarrow p \bar{p} e^{+} \nu_{e} $. No significant signal is
    observed, and an upper limit of $\mathcal{B}(D_{s}^{+} \rightarrow p \bar{p}
    e^{+} \nu_{e}) < 2.0 \times 10^{-4}$ is set at the 90\% confidence level.
    This measurement is useful input in  understanding the baryonic transition
    of $D_{s}^{+}$ mesons.
\end{abstract}


\maketitle

\oddsidemargin  -0.2cm    
\evensidemargin -0.2cm

\section{Introduction}
In the charm sector, probing the transition between charm meson and baryon pairs
is still largely an unexplored territory.  Phase-space constraints dictate that
only the $D_{s}^{+}$ meson can decay in such a manner. Until now, only one
baryonic mode, $D_{s}^{+} \to p \bar{n}$, has been observed.  It was first seen
by the CLEO Collaboration, with a branching fraction of $(1.30 \pm 0.4) \times
10^{-3}$~\cite{Athar:2008ug}, and subsequently confirmed by BESIII \cite{pnar}.
This mode is expected to be suppressed by chiral symmetry, and predictions for
its decay rate are several orders of magnitude below the observed
value~\cite{Pham:1980dc}, motivating the study of other baryonic channels.  A
promising candidate is the semileptonic decay mode $D_{s}^{+} \to p\bar{p} e^{+}
\nu_{e}$, for which theoretical calculations are expected to be more robust.
Recently, H. Y. Cheng and X. W. Kang \cite{Cheng:2017qpv} predicted a small
branching fraction, $\mathcal{B}(D_{s}^{+} \rightarrow p \bar{p} e^{+} \nu_{e})
\sim 10^{-8}$.   Even in this case, however, there are significant uncertainties
on the prediction, associated with the challenge of calculating the hadronic
form factor.   Experimental input is therefore needed to help illuminate this
poorly understood class of charm decays.

An additional motivation for searching for this decay is that the final state
provides an ideal laboratory to study near-threshold enhancement phenomenon.
This behavior was initially observed in the radiative process $J/\psi \to \gamma
p \bar{p}$ by BESIII~\cite{Bai:2003sw} and confirmed by
CLEO~\cite{Alexander:2010vd} and BESIII~\cite{BESIII:2011aa}, but not yet
observed in other processes~\cite{Athar:2005nu,Ablikim:2013cif,Ablikim:2015pkc}. 
A very attractive feature of searching for this phenomenon in $D_{s}^{+}
\rightarrow p \bar{p} e^{+} \nu_{e}$ decays is that the $p\bar{p}$ system is
produced close to mass threshold.

With the strong interaction dynamics described by a form factor $f_{+}(q^{2})$,
and in the limit of zero electron mass, the differential rate for the $D_{s}^{+}
\to p\bar{p} e^{+} \nu_{e}$ decay is given by
\begin{equation}
   \label{eq:decar rate}
    \frac{d \Gamma(D_{s}^{+} \to X e^{+} \nu_{e})}{d q^{2}}
    =  \frac{G_{F}^{2} |V_{cs}|^{2}} {24 \pi^{3}} p_{X}^{3}  |f_{+}(q^{2})|^{2},
\end{equation}
where $G_{F}$ is the Fermi constant, $V_{cs}$ is the Cabibbo-Kabayashi-Maskawa
(CKM) matrix element, the $X$ represents the $p\bar{p}$ system, which is assumed
to form a $^{1}S_{0}$ state, $p_{X}$ is the momentum of $p\bar{p}$ system in the
rest frame of the $D_{s}^{+}$ meson, and $q$ is the transition momentum between
$X$ and $D_{s}^{+}$.  The form factor $f_{+}(q^2)$ is described by the well
known ISGW2 model~\cite{Scora:1995ty}, 
\begin{equation}
    f_{+}(q^{2}) = f_{+}(q_{\rm max}^{2}) \left(1  + \frac{ r^{2} } 
     {12 }\left( q_{\rm max}^{2} - q^{2} \right)   \right)^{-1},
\end{equation}
where $r$ is the effective radius of the $D_{s}^{+}$ meson, and $q_{\rm
max}^{2}$ is the kinematic limit of $q^{2}$.

In this article,  we report a search for the decay $D_s^+ \to p \bar{p} e^+
\nu_e$ using a 3.19 fb$^{-1}$ data set collected at $\sqrt{s}=4.178$ GeV with
the BESIII detector operating at the BEPCII collider.  \section{BESIII DETECTOR
AND Monte carlo Simulation} The BESIII detector is a magnetic
spectrometer~\cite{Ablikim:2009aa} located at the Beijing Electron Position
Collider (BEPCII)~\cite{Yu:IPAC2016-TUYA01}. The cylindrical core of the BESIII
detector consists of a helium-based multilayer drift chamber (MDC), a plastic
scintillator time-of-flight system (TOF), and a CsI~(Tl) electromagnetic
calorimeter (EMC), which are all enclosed in a superconducting solenoidal magnet
providing a 1.0~T magnetic field. The solenoid
is supported by an octagonal flux-return yoke with resistive-plate counter
muon-identifier modules interleaved with steel. The acceptance of charged
particles and photons is 93\% over the $4 \pi$ solid angle. The charged-particle
momentum resolution at $1~{\rm GeV}/c$ is $0.5\%$, and the $dE/dx$ resolution is
$6\%$ for the electrons from Bhabha scattering. The EMC measures photon energies
with a resolution of $2.5\%$ ($5\%$) at $1$~GeV in the barrel (end-cap) region.
The time resolution of the TOF barrel part is 68~ps. The end-cap TOF system was
upgraded in 2015 with multi-gap resistive plate chamber technology, providing a
time resolution of 60 ps~\cite{Li2017, Sun2017}.

Simulated events are generated with a {\sc{geant4}}-based~\cite{geant4} software
package using a detailed description of the detector geometry and of the
particle interactions in the detector material. A sample of inclusive Monte
Carlo (MC) simulation is produced at $\sqrt{s}=4.178$~GeV. This sample includes
all known open-charm decay processes and the $c\bar{c}$ resonances, $J/\psi$,
$\psi(3686)$ and $\psi(3770)$ via the initial state radiation (ISR).
Additionally, the continuum process ($e^{+}e^{-}\to q\bar{q}$, $q=u$, $d$, and
$s$), Bhabha scattering, $\mu^+\mu^-$, $\tau^+\tau^-$, as well as two-photon
process are included. The open charm processes are generated using
{\sc{conexc}}~\cite{Ping:2013jka} and their subsequent decays are modeled by
{\sc{evtgen}}~\cite{Ping:2008zz} with the known branching fractions from the
Particle Data Group~\cite{Patrignani:2016xqp}, and the remaining unknown decay
modes of the narrow $c\bar{c}$ resonances are generated using the modified
{\sc{lund}} model~\cite{ref:lundcharm}. The signal model is described by Eq.
\ref{eq:decar rate}. We assume that the $p\bar{p}$ S-wave system dominates in
the decay and adopt a non-resonance S-wave to describe the $p\bar{p}$ system
(when assigning the systematic uncertainties we also consider the possibility of
a resonance contributing to the decay).
\section{Analysis Method}
Throughout the paper, charge-conjugate modes are implicitly implied, unless
otherwise noted. The $D_{s}^{\pm}D_{s}^{*\mp}$ pairs are produced at a
center-of-mass energy of 4.178~GeV. The double tag (DT) method is employed to
perform a measurement  of the absolute branching fraction. We first select
``single tag" (ST) events in which either a $D_s^-$ or $D_s^+$ meson is fully
reconstructed. Then the $D_s^{+}$ decay of the interest is searched  for in the
remainder of each event, namely, in DT events where both the $D_s^+$ and $D_s^-$
are fully reconstructed, regardless of the $\gamma$ or $\pi^{0}$ emitted from
$D_s^{*\pm}$ meson. The absolute branching fraction for the $D_s^+$ meson decay
is calculated for each tag mode $\alpha$, and is given by
\begin{equation}
     \mathcal{B}_{\rm sig}^{\alpha} = \frac{ N_{\rm DT}^{\alpha} }
     {N_{\rm ST}^{\alpha} \epsilon_{\rm DT}^{\alpha}
     /\epsilon_{\rm ST}^{\alpha}},
 \end{equation}
where $N_{\rm ST}^{\alpha}$ and $N_{\rm DT}^{\alpha}$ are the yields of ST
events and DT events, respectively, and $\epsilon_{\rm ST}^{\alpha}$ and
$\epsilon_{\rm DT}^{\alpha}$ are the ST and DT efficiencies for the tag mode
$\alpha$.  \subsection{ST Analysis} All charged tracks must have a polar angle
($\theta$) within $|\cos\theta|<$ 0.93, where $\theta$ is measured with respect
to the direction of the beam. Furthermore, all charged tracks, apart from those
from $K^0_S$ candidates, are required to point back to the interaction
point~(IP).  This is achieved  by imposing  $V_{r}<1$~cm and $|V_{z}|<10$~cm,
where $V_{r}$ and $|V_{z}|$ are the distances of the closest approach to the IP
in the transverse plane and along the positron beam direction, respectively.
The information from the $dE/dx$ and TOF measurements are combined to evaluate
the particle identification (PID) probability ($\mathcal{L}$). A charged track
is assigned to be a kaon~(pion) candidate if it satisfies $ {\mathcal
L}_{K(\pi)}>{\mathcal L}_{\pi(K)}$.  Candidate $K_S^0$ mesons are formed from
two oppositely charged tracks satisfying $|V_{z}|<20$~cm and $|\cos
\theta|<0.93$, which are assumed to be pions without the imposition of further
PID requirements. These two tracks are  constrained to have a common vertex and
the invariant mass of the  pair is required to lie within
(0.487,\,0.511)~GeV/$c^2$. The decay length of the $K^0_S$ candidates is
required to be larger than twice the uncertainty of the decay length.

The $D_{s}^{-}$ single-tag candidates are reconstructed in the three tag modes,
$K^+ K^- \pi^-$, $K_S^0 K^-$ and $K_S^0 K^+ \pi^- \pi^-$, which all have high
signal-to-noise ratios and yield the highest sensitivity, according to studies
performed on the inclusive MC sample.

To suppress the background involving $D^{*} \to D \pi$ decays, the momenta of
pions from the $D_s^-$ decay are required to be greater than 0.1 GeV/$c$. The
recoil mass evaluated against the $D_{s}^{-}$ candidate, $M_{\rm
recoil}(D_{s}^{-}) = \sqrt{\left( \sqrt{s} - E_{ D_{s}^{-} }\right)^{2} -
|\vec{p}_{D_{s}^{-}}|^{2}}$, is used  to reject background from non-$D_s^\pm
D_s^{*\mp}$ processes with the requirement that $2.06 < M_{\rm
recoil}(D_{s}^{-}) < 2.18~{\rm GeV}/c^2 $.  If there are several  $D_{s}^{-}$
candidates in the event, only that one with recoil mass closest to the
$D_{s}^{*+}$ nominal mass is retained. 

An unbinned maximum-likelihood fit is performed on the $M_{D_s^-}$ spectrum of
each of the three selected ST tag modes, as shown in Fig.~\ref{Stag yield of
data}. In the fit, the signal shape is taken from the distribution found in MC
simulation, using the kernel-estimation method~\cite{Cranmer:2000du} provided as
a RooKeysPdf class in ROOT~\cite{ROOT}, convolved with a Gaussian function. The
non-peaking background  is described by a second- or third-order Chebyshev
polynomial. The small peaking contribution seen in the $D_s^-\to K^0_S\pi^-$
mode is from $D^{-} \to K_{S}^{0}\pi^{-}$ decays and its shape is taken from MC
simulation, with the absolute normalization determined from the fit.

\begin{figure}[htbp] 
    \begin{overpic} [width = 0.9\linewidth] {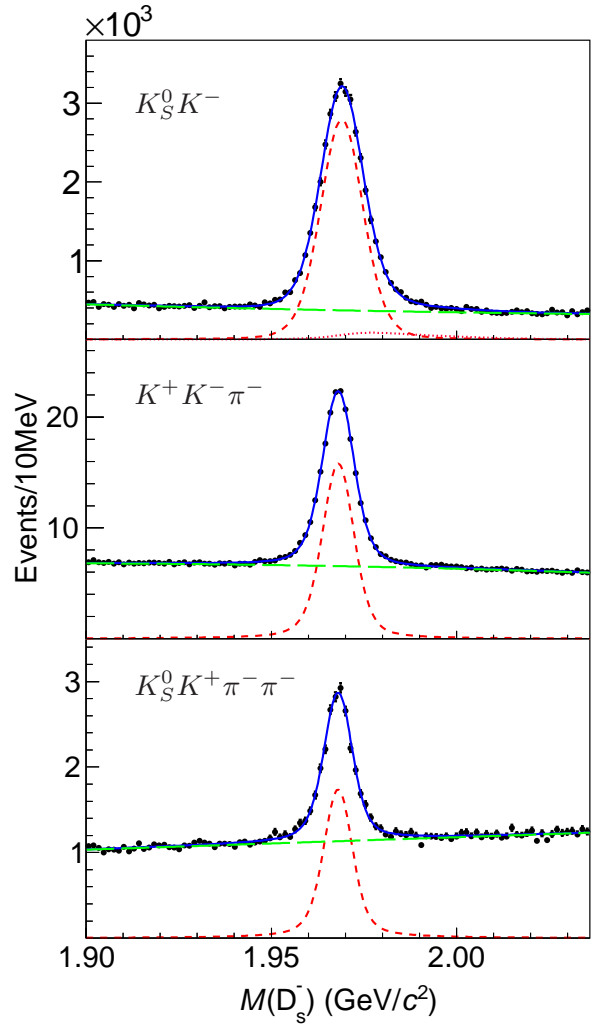}
        \put(12, 61){{\scalebox{1.2}{$K^{+} K^{-}\pi^{-}$}} }
        \put(12, 89){{\scalebox{1.2}{$K_{S}^{0} K^{-}$}} }
        \put(12, 33){{\scalebox{1.2}{$K_{S}^{0} K^{+} \pi^{-} \pi^{-}$}} }
    \end{overpic}

    \caption{
        Fit to the $M_{D_s^-}$ spectrum for each tag mode. The dots with error
        bars are from the data. The blue solid lines represent the total fit
        result. The red dashed line and green long and dashed line are the
        signal shape, and non-peaking background. The pink dotted line in the
        $K_{S}^{0} K^{-}$ tag mode corresponds to the peak background due to
        $D^{-} \to K_{S}^{0} \pi^{-}$.
    }
    \label{Stag yield of data}
\end{figure}

All the selected $D_{s}^{-}$ candidates are retained for further analysis. The
resultant yields, $N_{\rm ST}^{\alpha}$, and the corresponding selection
efficiencies $\epsilon_{\rm ST}^{\alpha}$, as determined from the simulation,
are summarized in Table~\ref{tab:tag_yields}. The total yield of single tags is
$N^{\rm tot}_{\rm ST} = 186091 \pm 719$, where the uncertainty is statistical.

\begin{table}[htbp]
    \caption{
        Summary of $N_{\rm ST}^{\alpha}$, $\epsilon_{\rm ST}^{\alpha}$ and
        $\epsilon_{\rm DT}^{\alpha}$ for the tag mode $\alpha$. All
        uncertainties are statistical only.
    }
    \centering
    \begin{tabular}{l r@{$\pm$}c r@{$\pm$}c r@{$\pm$} c} 
        \hline\hline
        Mode &\multicolumn{2}{c}{$N_{\rm ST}^{\alpha}$} 
        & \multicolumn{2}{c}{$\epsilon_{\rm ST}^{\alpha}$~(\%)} 
        & \multicolumn{2}{c}{$\epsilon_{\rm DT}^{\alpha}$~(\%)}   \\ \hline
        $K^0_SK^+$		& 31267 &261& 42.32  & 0.04 & 8.63 &  0.07 \\
        $K^+K^-\pi^+$	&140277 &635&  49.33  & 0.18 & 9.62 &  0.08 \\	
        $K^0_SK^-\pi^+\pi^+$	& 14547 & 214& 21.08  & 0.07 &  3.82  & 0.04 \\
        \hline\hline
     \end{tabular}
    \label{tab:tag_yields}
\end{table}
\subsection{DT Analysis}
After the reconstruction of the ST $D_{s}^{-}$ candidate,  there are required to
be fewer than four unused charged tracks in the event.  We search for proton and
electron candidates among these unused tracks. The charged tracks are assigned
as proton candidates if they satisfy ${\mathcal L}_{p}>{\mathcal L}_{K}$, and
${\mathcal L}_{p}>{\mathcal L}_{\pi}$.

\begin{figure}[htpb]
    \centering
    \subfigure{
        \begin{overpic} [width = 0.9\linewidth] {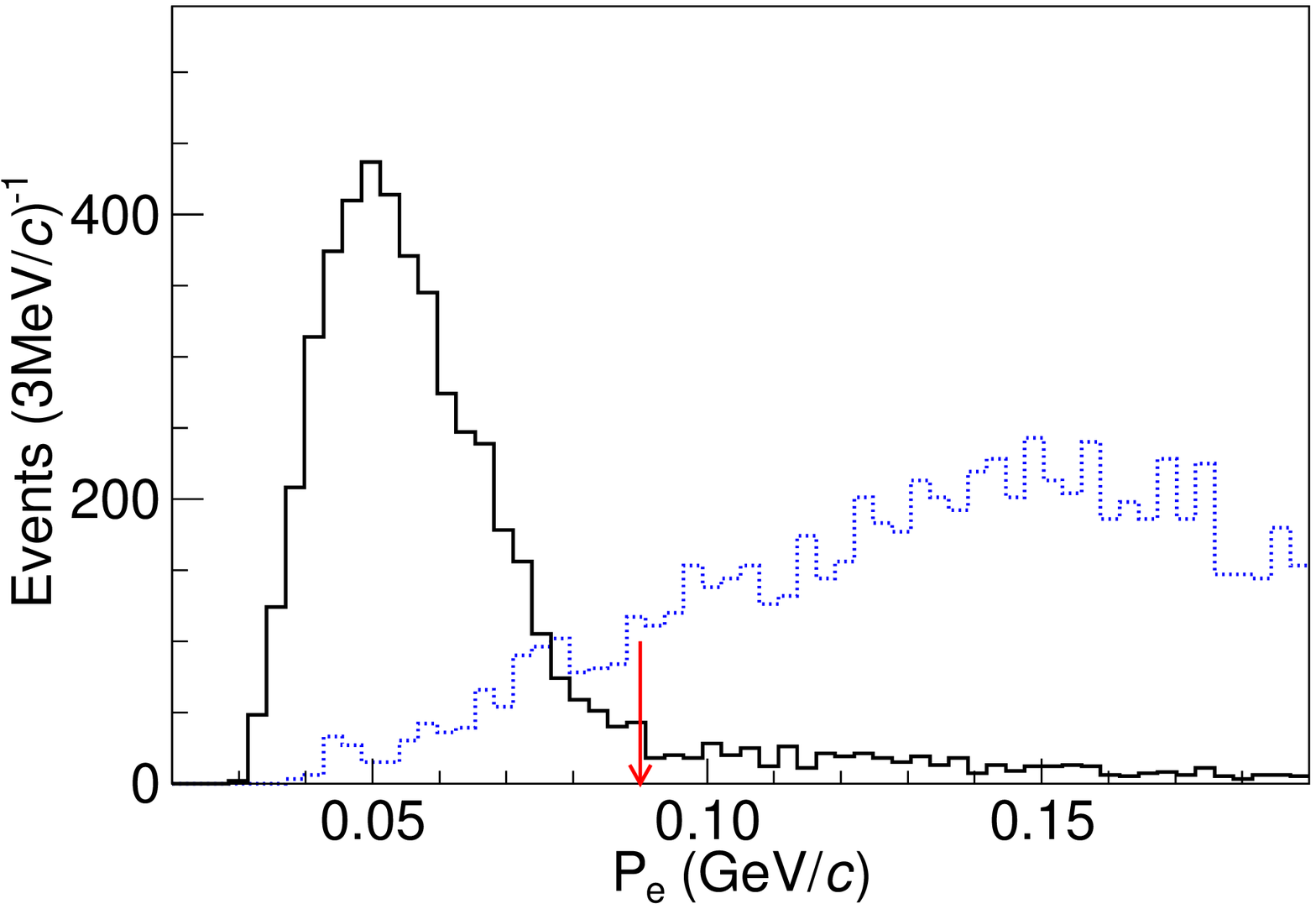}
             \put (73 , 55){  
             {\scalebox{1.5}{$(a)$ }} }
        \end{overpic}
        \label{fig:pele}
    }
    \subfigure{
        \begin{overpic} [width = 0.9\linewidth]{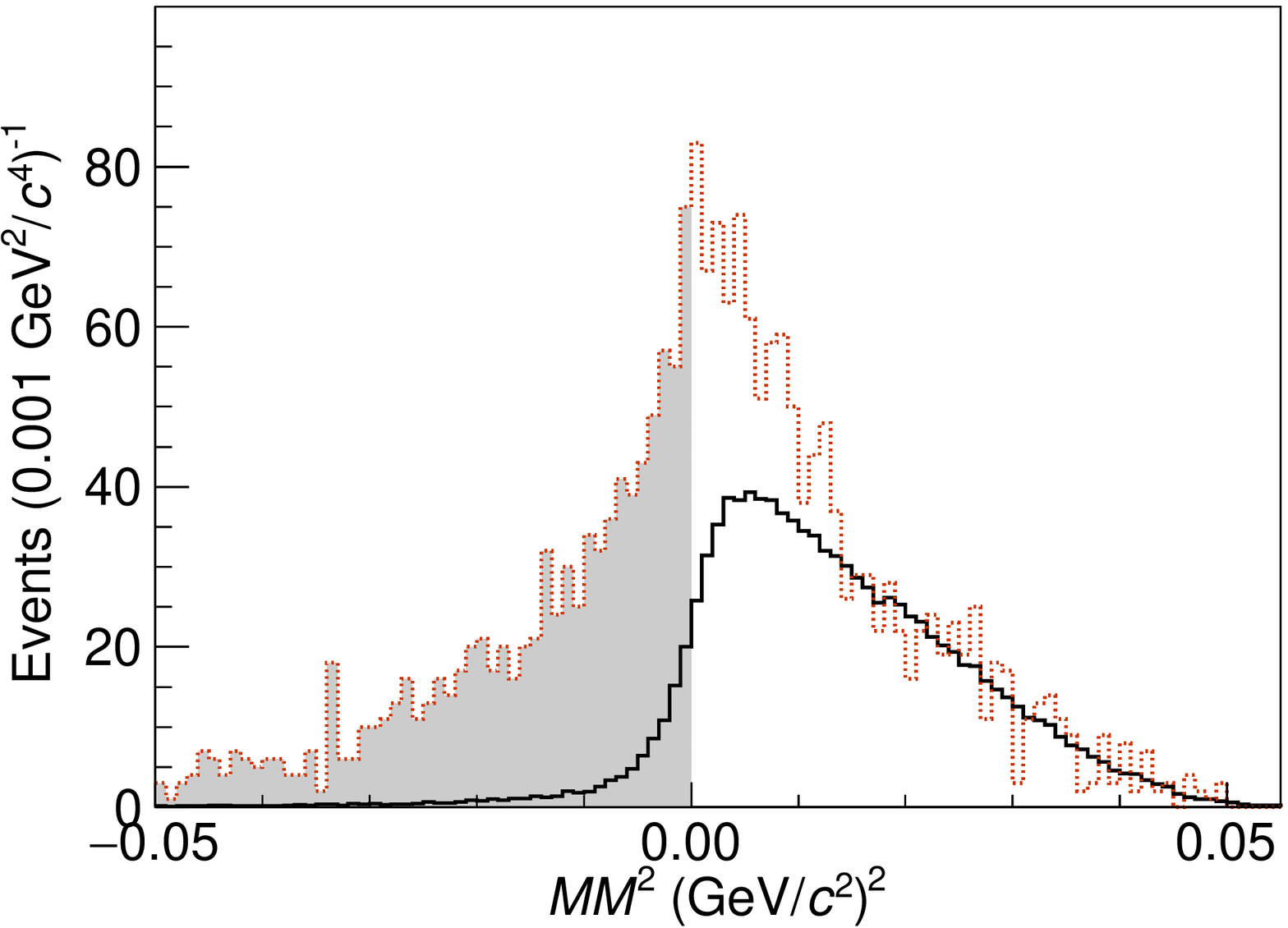}
             \put (73 , 55 ){ 
             {\scalebox{1.5} {$(b)$} } }
        \end{overpic}
    \label{fig:MM}
    }
    \caption{
        (a) The electron momentum ($P_{e}$) distribution in MC simulation. The
        green and blue solid histograms represent the signal and
        background distribution, respectively. The red arrow shows
        the maximum value of $P_{e}$ allowed in the selection. (b) The distribution of
        $MM^{2}$ from MC simulation. The red dotted histogram shows the
        background distribution from the inclusive MC sample, which is
        completely dominated by the continuum $q\bar{q}$ process.  The grey region is
        rejected by the requirement that
        $MM^{2} > 0\ {\rm GeV}^{2}/c^{4}$.
        The solid green histogram shows the signal distribution.
    }
\end{figure}

As shown in  Fig.~\ref{fig:pele} the momentum of the electron in the signal
decay is typically very low.  Consequently, in most decays the electron is not
reconstructed in the detector.  The presence of the $p\bar{p}$ pair, however, is
a sufficiently distinctive signature for such events to be classified as signal,
even in the cases when there is no track reconstructed corresponding to the
electron.  In those cases when a third track is found with lower momentum than
the $p$ and $\bar{p}$ candidates, which happens in about 5\% of selected events,
this track is assigned to be  the electron candidate without any PID
requirement.  Requiring that the momentum of the electron candidate is smaller
than 0.09 GeV/$c$ reduces background, whose spectrum is also shown in
Fig.~\ref{fig:pele}.  The missing mass-squared $MM^{2} = \left(\sqrt{s} - E_{\rm
tag} - E_{\rm sig}\right)^{2} - \left(\vec{p}_{\rm tag} + \vec{p}_{\rm
sig}\right)^{2}$ is required to be larger than $0 $ GeV$^{2}/c^{4}$ to further
reduce the background from continuum $q\bar{q}$ production, as shown in
Fig.~\ref{fig:MM}.  Here $E_{\rm tag}$, $\vec{p}_{\rm tag}$ and $E_{\rm sig}$,
$\vec{p}_{\rm sig}$ are the total energy and momentum of the tag side and signal
side, respectively. As we ignore the momentum of the photon or $\pi^{0}$ from
the $D_{s}^{*}$ decay, the signal has a predominantly positive value  of
$MM^{2}$ as can be seen in Fig.~\ref{fig:MM}. The DT efficiencies $\epsilon_{\rm
DT}^{\alpha}$ as summarized in Table~\ref{tab:tag_yields} are determined from
simulation and later corrected for the tracking and PID differences between data
and MC simulation.

An extended unbinned maximum-likelihood fit to the $M_{D_s^-}$ distribution of
the tag meson is used to determine the number of DT signal events. For the tag
mode $\alpha$, the likelihood value is defined as
\begin{eqnarray}
    \mathcal{L}^{\alpha} =&& \frac{e^{- (N_{\rm sig}^{\alpha} + N_{\rm bkg}^{\alpha})}}{n^{\alpha} !}\times \\ \notag
    &&
     \prod_{i=1}^{n^{\alpha}}
    \left( N_{\rm sig}^{\alpha} \mathcal{P}_{\rm sig}^{\alpha}(M_{D_{s}^{-}}) +N_{\rm bkg}^{\alpha}\mathcal{P}_{\rm bkg}^{\alpha}(M_{D_{s}^{-}}) \right),
\end{eqnarray}
where $n^{\alpha} = N_{\rm sig}^{\alpha} + N_{\rm bkg}^{\alpha}$ is the number
of total observed DT events. $N_{\rm sig}^{\alpha}$ and $N_{\rm bkg}^{\alpha}$
denote the fitted yields for signal and backgrounds, respectively, and
$\mathcal{P}_{\rm sig}^{\alpha}$ and $\mathcal{P}_{\rm bkg}^{\alpha}$ are the
corresponding probability density functions (PDF) in the fit.  The PDF
distributions are taken from simulation, with the inclusive MC sample being used
to represent the background.

A simultaneous fit to the $M_{D_s^-}$ spectra from the three tag modes is
performed with the combined likelihood $\mathcal{L}^{\rm com} =
\prod_{\alpha=1}^{3} \mathcal{L}^{\alpha}$, sharing the same branching fraction
of $D_s^+ \to p \bar{p} e^+ \nu_e$ for each.

\begin{figure}[htbp]
    \mbox{
        \begin{overpic} [width = 0.9\linewidth]{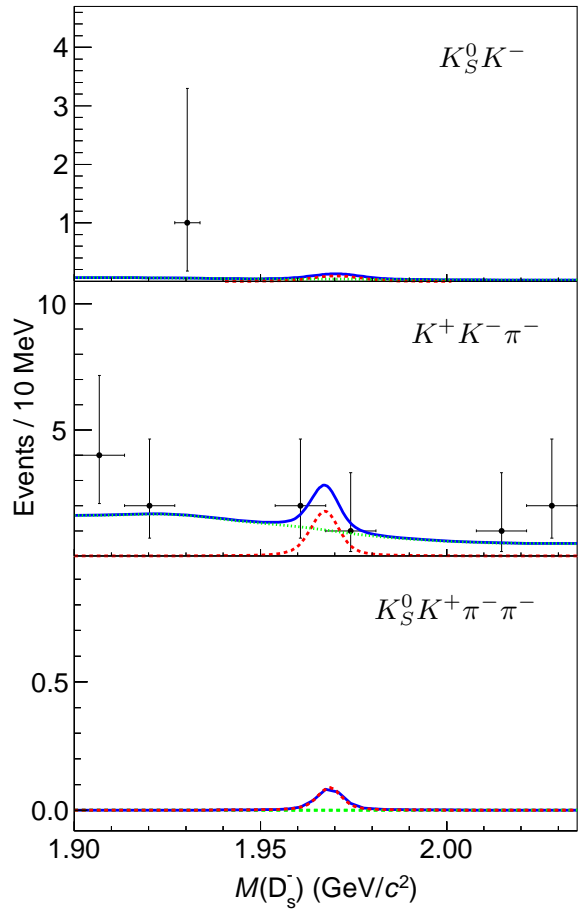}
            \put (44, 63){ \color {black} {\scalebox{1.2}{$K^{+} K^{-}\pi^{-}$}} }
             \put (47, 93){ \color {black} {\scalebox{1.2}{$K_{S}^{0} K^{-}$}} }
             \put (40, 33 ){ \color {black} {\scalebox{1.2}{$K_{S}^{0} K^{+} \pi^{-} \pi^{-}$}} }
        \end{overpic}
    }
      \caption{
          Fits to $M_{D_s^-}$ after DT event selection. The points with error bars
          are data, and the blue solid lines show the total fit result. The red
          and green dotted lines denote the signal and background shapes,
          respectively.
      }
      \label{fig:DT Yield}
\end{figure}

The fit results are shown in Fig.~\ref{fig:DT Yield}. The signal yields for the
three selected ST modes are determined to be $0.3 ^{+0.4}_{-0.3}$,
$1.4^{+1.8}_{-1.3}$, $0.1 \pm  0.1$, respectively, and the branching fraction is
measured to be $ \mathcal{B}(D_{s}^{+} \to p\bar{p} e^{+}\nu_{e}) = (0.50
^{+0.63}_{-0.44} )\times 10^{-4}$ with a significance of $1.2\sigma$, where the
uncertainty is statistical.  Since no significant signals are seen, we set an
upper limit after taking into account the systematic uncertainties.

\section{Systematic Uncertainties}
\label{sec:background}

Possible sources of systematic bias are investigated, and corresponding
uncertainties are assigned as discussed below.  These uncertainties are listed
and added in quadrature in Table~\ref{Tab: uncertainty}, apart from that
component associated with the fit of the DT yields, which is accounted
separately.

\subsection{Fitting ST ${\mathbf{M(D_s^-)}}$ yields}
A set of alternative fits is performed, in which the following variations are
applied: the background shape is changed from a second-  to  a third-order
Chebyshev polynomial;  the signal shape is changed from the MC-simulated shape
convolved with a single Gaussian function to the sum of two Gaussian functions;
and the fitting range is both increased and decreased  by 5~MeV$/c^{2}$.  The
procedures are performed both on inclusive MC and data, and the overall sum in
quadrature of the observed differences in  the efficiency corrected signal
yields is taken as the systematic uncertainty associated with fitting the ST
$M_{D_s^-}$ yields.

\subsection{Tracking and PID}
The uncertainties associated with the knowledge of the tracking and PID
efficiencies for the proton and anti-proton are studied with a control sample of
$e^{+}e^{-} \to p \bar{p} \pi^{+} \pi^{-}$ decays. The signal efficiency is
re-weighted according to the momentum distributions of the proton  and
anti-proton. The uncertainties associated with the tracking and PID efficiencies
are assigned to be 2.9\% and 2.2\%, respectively.

\subsection{${\mathbf{MM^{2}}}$ requirement}
The systematic uncertainty from the $MM^{2}$ requirement is associated with the
knowledge of the detector resolution. To estimate this uncertainty a control
sample is selected, which has the same tag modes for the $D_s^-$ as in the
nominal analysis, and where the  other meson is reconstructed in the mode
$D_{s}^{+} \to K^{+} K^{-} \pi^{+}$, with the pion then removed and treated  as
a missing particle.  The $MM^{2}$ resolution is compared between data and MC
simulation, and the difference is applied as additional smearing to the signal
MC sample.  The difference between the selection efficiencies with this
treatment and the nominal analysis is assigned as the systematic uncertainty due
to the $MM^2$ requirement.

\subsection{MC modeling}
To estimate the systematic uncertainty due to the possibility of a $p\bar{p}$
bound state, and its assumed mass and width, we simulate and analyze new MC
samples that include a resonant system in the decay.  We vary the mass of the
system from 1.80 to 1.85~GeV/$c^2$, and the width from 10 to 100
MeV/$c^2$~\cite{Bai:2003sw,Alexander:2010vd,BESIII:2011aa}. The largest relative
change of the signal efficiency is found to be 18\% and is assigned as the
uncertainty from MC modeling. 

\subsection{Fitting}
It is only necessary to consider the uncertainty on the knowledge of the
background shape, as that associated with the signal distribution has negligible
impact on the result.  The background shape is obtained using the kernel
estimation method \cite{Cranmer:2000du} provided as a RooKeysPdf Class in
ROOT~\cite{ROOT}, based on the inclusive MC sample. Unlike the other sources of
uncertainties, the background shape affects the likelihood function directly. We
vary the smoothing parameter of RooKeysPdf within a reasonable range to obtain
alternative background shapes.  We adopt the background shape that gives the
largest upper limit on the signal branching ratio to assign the value of this
component of the systematic uncertainties.

\begin{table}[htbp]
    \caption{The relative systematic uncertainties (in percent).}
    \label{Tab: uncertainty}
    \centering
    \begin{tabular}{p{4.0cm}m{2.6cm}<{\centering}}
        \hline \hline
        Source            & Uncertainty \\ \hline
        ST yields         & 0.8   \\
        Tracking efficiency            & 2.9     \\
        PID efficiency                 & 2.2    \\
        $MM^{2}$ requirement           & 1.0     \\
        MC modeling                    & 18   \\
        \hline
        Total                          & 19 \\
        \hline \hline
    \end{tabular}
\end{table}

\section{Result and Summary}

The upper limit (UL) on the branching fraction is set at the 90\% confidence level
(CL) according to
\begin{equation}
    \label{eq:UL}
    \frac{ \int_{0}^{\rm UL} L(\mathcal{B}) d \mathcal{B} }
    {\int_{0}^{1} L(\mathcal{B}) d \mathcal{B}} = 0.9.
\end{equation}
\begin{figure}[htbp]
    \begin{overpic}[width = 0.4\textwidth]{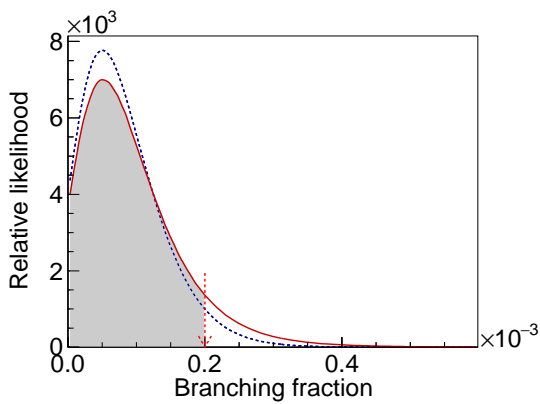}
    \end{overpic}
 
    \caption{
        The likelihood distribution. The blue dotted line denotes the likelihood
        distribution before the smearing, while the red solid line shows the
        smeared likelihood.
    }
\label{fig:prob}
\end{figure}

Taking the systematic uncertainties ($\sigma_{\epsilon}$)~into account
\cite{K.Stenson:2006}, the likelihood distribution of the branching fraction,
$L(\mathcal{B})$ is determined by
\begin{equation}
    \label{eq:smear}
    L(\mathcal{B})\varpropto \int _{0} ^{1} L^\prime(
    \frac{\epsilon}{\epsilon_{0}}\mathcal{B}) \
    e^{  -\frac{(\epsilon -  \epsilon_{0})^{2}}
        {2 \sigma_{\epsilon}^{2}}
    } d \epsilon,
\end{equation}
where $L^\prime$ denotes the likelihood of the fit result, $\epsilon_{0}$ is the
nominal signal efficiency based on the signal MC sample, and $\sigma_{\epsilon}$
is the systematic uncertainty associated with the signal efficiency. The
likelihood $L^\prime$ and smeared likelihood $L$ distributions are shown in
Fig.~\ref{fig:prob}, and the UL is denoted by the red arrow. 

In summary, by analyzing 3.19 fb$^{-1}$ of $e^{+}e^{-}$ annihilation sample
collected at $\sqrt{s} = 4.178 $ GeV with the BESIII detector, we perform the
first search on the decay $D_{s}^{+} \to p \bar{p} e^{+} \nu_{e}$, and an upper
limit is set at the 90\% CL of
\begin{equation}
   \notag
    \mathcal{B}(D_{s}^{+} \to p\bar{p} e^{+}\nu_{e})  < 2.0 \times 10^{-4}.
\end{equation}
In order to improve this limit, and approach the predicted branching ratio of
Ref.~\cite{Cheng:2017qpv}, larger data samples are needed, either at BESIII or
at future experiments such as Belle II experiment~\cite{Kou:2018nap} and super
tau-charm factory~\cite{Bondar:2013cja,Zhou:2016qfu}.

\vspace{1cm}

\begin{acknowledgments}
The BESIII collaboration thanks the staff of BEPCII and the IHEP computing center for their strong support. This work is supported in part by National Key Basic Research Program of China under Contract No.~2015CB856700;
National Natural Science Foundation of China (NSFC) under Contracts Nos.
    11235011, 11335008, 11425524, 11625523, 11635010, 11935018;
the Chinese Academy of Sciences (CAS) Large-Scale Scientific Facility Program; the CAS Center for
Excellence in Particle Physics (CCEPP); Joint Large-Scale Scientific Facility Funds of the NSFC and CAS under Contracts Nos. U1332201, U1532257, U1532258; CAS Key Research Program of Frontier Sciences under Contracts Nos.
QYZDJ-SSW-SLH003, QYZDJ-SSW-SLH040; 100 Talents Program of CAS; National 1000 Talents Program of China; INPAC and Shanghai Key Laboratory for Particle Physics and Cosmology; German Research Foundation DFG under Contracts
Nos. Collaborative Research Center CRC 1044, FOR 2359; Istituto Nazionale di Fisica Nucleare, Italy; Koninklijke Nederlandse Akademie van Wetenschappen (KNAW) under Contract No.~530-4CDP03; Ministry of Development of
Turkey under Contract No.~DPT2006K-120470; National Natural Science Foundation of China (NSFC) under Contracts Nos. 11505034, 11575077; National Science and Technology fund; The Swedish Research Council; U. S. Department
of Energy under Contracts Nos.~DE-FG02-05ER41374, DE-SC-0010118, DE-SC-0010504, DE-SC-0012069; University of Groningen (RuG) and the Helmholtzzentrum fuer Schwerionenforschung GmbH (GSI), Darmstadt; WCU Program of National
Research Foundation of Korea under Contract No. R32-2008-000-10155-0.
\end{acknowledgments}

\end{document}